\documentclass[superscriptaddress,
twocolumn,showpacs,preprintnumbers,amsmath,amssymb]{revtex4}
\usepackage{graphicx}
\usepackage{dcolumn}
\usepackage{bm}
\usepackage{latexsym}
\begin{document}
\title{Traffic of interacting ribosomes on mRNA during protein synthesis:\\ 
effects of chemo-mechanics of individual ribosomes
}
\author{Aakash Basu}
\affiliation{Department of Physics, Indian Institute of Technology,
Kanpur 208016, India.}
\author{Debashish Chowdhury{\footnote{Corresponding author(E-mail: debch@iitk.ac.in)}}}
\affiliation{Department of Physics, Indian Institute of Technology,
Kanpur 208016, India.}
\date{\today}%
\begin{abstract}
Many {\it ribosomes} simultaneously move on the same messenger RNA
(mRNA), each synthesizing separately a copy of the same protein. 
In contrast to the earlier models, here {\it we develop a ``unified'' 
theoretical model} that not only incorporates the {\it mutual 
exclusions} of the interacting ribosomes, but also describes explicitly 
the mechano-chemistry of each of these macromolecular machines during 
protein synthesis. Using analytical and numerical techniques of 
non-equilibrium statistical mechanics, we analyze the rates of 
protein synthesis and the spatio-temporal organization of the 
ribosomes in this model. We also predict how these properties would 
change with the changes in the rates of the various chemo-mechanical 
processes in each ribosome. Finally, we illustrate the power of this 
model by making experimentally testable predictions on the rates of 
protein synthesis and the density profiles of the ribosomes on some 
mRNAs in {\it E-coli}.

\end{abstract}
\maketitle
\section{Introduction}

Synthesis of each protein from the corresponding messenger RNA (mRNA) 
is carried out by a ribosome \cite{spirinbook} and the process is  
referred to as {\it translation} (of genetic code). Ribosome is one 
of the largest and most sophisticated {\it macromolecular machines} 
within the cell \cite{spirin02,frank06}. It is not merely a 
``protein-making motor protein'' \cite{cross97,hill69} but it also 
serves as a ``workshop'' which provides a platform where a coordinated 
action of many tools take place for the synthesis of each of the 
proteins. What makes it very interesting from the perspective of 
statistical physics is that most often many ribosomes move 
simultaneously on a single mRNA strand while each synthesizes a 
separate copy of the same protein. Such a collective movement of the 
ribosomes on a single mRNA strand has superficial similarities with 
vehicular traffic \cite{css} and, therefore, will be referred to as 
ribosome traffic \cite{polrev,tgf06,physica}. All the earlier works  
\cite{macdonald68,macdonald69,lakatos03,shaw03,shaw04a,shaw04b,chou03,chou04,dong} 
treat ribosome traffic as a problem of non-equilibrium statistical 
mechanics of a system of interacting ``self-driven'' hard rods. 
But, strictly speaking, a ribosome is neither a particle nor a hard 
rod; its mechanical movement along the mRNA track is coupled to its 
internal mechano-chemical processes which drive the synthesis of the 
protein. Thus, one serious limitation of the earlier models is that 
these cannot account for the effects of the intra-ribosome chemical 
and conformational transitions on their collective spatio-temporal 
organization.

In this paper we develop a model that not only incorporates the 
inter-ribosome steric interactions (mutual exclusion), but also 
captures explicitly the essential steps in the intra-ribosome 
chemo-mechanical processes. From a physicist's perspective, our 
model is a biologically motivated extension of some earlier models 
developed for studying the collective spatio-temporal organization 
in a non-equilibrium system of interacting ``self-driven'' hard rods. 
However, in contrast to the earlier models, each of the rods in 
our model has several ``internal'' states which capture the 
different chemical and conformational states of an individual 
ribosome during its biochemical cycle.

Our modelling strategy for incorporating the biochemical cycle of 
the ribosomes is similar to that followed in the recent work 
\cite{nosc} on single-headed kinesin motors KIF1A. However, the 
implementation of the strategy is more difficult here not only 
because of the higher complexity of composition, architecture and 
mechano-chemical processes of the ribosomal machinery but also 
because of the sequence {\it heterogeneity} of the mRNA track 
\cite{nelson1,nelson2}. Our approach is based on a stochastic 
chemical kinetic model that describes the dynamics in terms of 
master equations. But our model makes no commitments to either power 
stroke or Brownian ratchet mechanism \cite{julicher,reimann,howard06} 
of molecular motors.

The paper is organized as follows: Because of the interdisciplinary 
nature of the topic investigated in this paper, we present in section 
\ref{bio} a summary of the essential biochemical and mechanical 
processes during a complete operational cycle of a single ribosome.
We present a brief critical review of the earlier models in section 
\ref{oldmodels} followed by a description of our model in section 
\ref{model} so as to highlight the novel features of our model. 
We report our results on this model with periodic boundary conditions 
(PBC) in section \ref{extended} and those with open 
boundary conditions (OBC) in section \ref{openbc}. We summarize the main 
results and draw conclusions in section \ref{conclusion}.

\section{Summary of the essential chemo-mechanical processes}\label{bio}

A protein is a linear polymer of amino acids which are linked together 
by peptide bonds and, therefore, often referred to as a polypeptide. 
An mRNA is a linear polymer of nucleotides and triplets of nucleotides 
form one single codon. The stretch of mRNA between a start codon and 
a stop codon serves as a template for the synthesis of a polypeptide. 
The process of translation itself can be divided into three main 
stages: (a) {\it initiation}, during which the ribosomal subunits 
assemble on the start codon on the mRNA strand, (b) {\it 
elongation}, during which the nascent polypeptide gets elongated by 
the formation of peptide bonds with new amino acids, and 
(c) {\it termination}, during which the process of translation gets 
terminated at the stop codon and the polypeptide is released. 
Initiation or termination can be the rate-limiting stage in the 
synthesis of a protein from the mRNA template \cite{shaw04a}.
In this paper we shall be concerned mostly with the process of 
{\it elongation}. 

The specific sequence of amino acids on a polypeptide is dictated  
by the sequence of codons on the corresponding template mRNA. 
The dictionary of translation relates each type of possible codons 
with one of the 20 species of amino acids; these correspondences are 
implemented by a class of adaptor molecules called tRNA. One end of 
a tRNA molecule consists of an anti-codon (a triplet of nucleotides) 
while the other end carries the cognate amino acid (i.e., the amino 
acid that corresponds to its anti-codon). Since each species of 
anti-codon is exactly complimentary to a particular species of codon, 
each codon on the mRNA gets translated into a particular species of 
amino acid on the polypeptide. A tRNA molecule bound to its cognate 
amino acid is called aminoacyl-tRNA (aa-tRNA).

\begin{figure}[t] 
\begin{center}
\includegraphics[width=0.9\columnwidth]{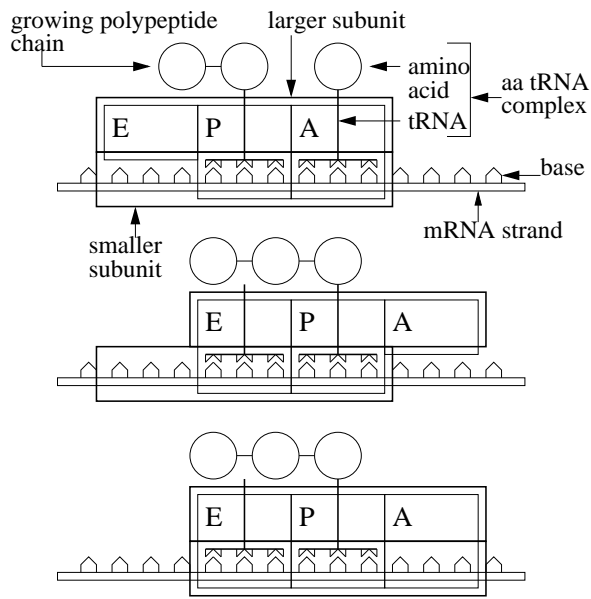}
\end{center}
\caption{A pictoral depiction of three major steps in the 
chemo-mechanical cycle of a single ribosome. The larger and 
smaller subunits have been depicted as two rectangles. The 
complementary shapes of the vertical tips and dips merely 
emphasize the codon-anticodon matching.  }  
\label{ribocartoon}
\end{figure}

\begin{figure}[t] 
\begin{center}
\includegraphics[width=0.9\columnwidth]{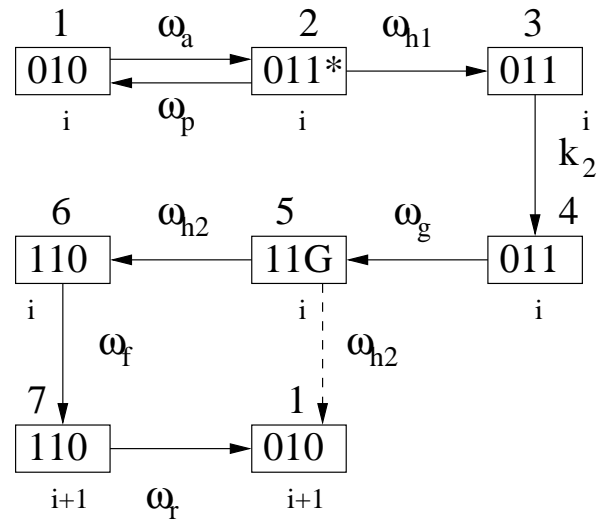}
\end{center}
\caption{A schematic representation of the biochemical cycle of a 
single ribosome during the elongation stage of translation in our 
model. Each box represents a distinct state of the ribosome. The 
index below the box labels the codon on the mRNA with which the 
smaller subunit of the ribosome binds. The number above the box 
labels the biochemical state of the ribosome. Within each box, 
$1 (0)$ represents presence (absence) of tRNA on binding sites 
E, P, A respectively. $1^{*}$ is a elongation factor (EF)-Tu bound 
tRNA and G is a EF-G GTPase. The symbols accompanied by the arrows 
define the rate constants for the transitions from one biochemical 
state to another. As explained in section \ref{model}, the dashed 
arrow represents the approximate pathway we have considered in the 
simplified dynamics of our model.}  
\label{states}
\end{figure}

Each ribosome consists of two parts which are usually referred to as 
the larger and the smaller subunits. There are four binding sites on 
each ribosome. Of these, three sites (called E, P, A), which are 
located in the larger subunit, bind to aminoacyl-tRNA (aa-tRNA), while 
the fourth binding site, which is located on the smaller subunit, 
binds to the mRNA strand. The translocation of the smaller subunit 
of each ribosome on the mRNA track is coupled to the biochemical 
processes occuring in the larger subunit. 

Three major steps in the biochemical cycle of a ribosome are sketched 
schematically in fig.\ref{ribocartoon}. In the first, the ribosome 
selects an aa-tRNA whose anticodon is exactly complementary to the codon 
on the mRNA. Next, it catalyzes the reaction responsible for the 
formation of the peptide bond between the existing polypeptide and the 
newly recruited amino acid resulting in the elongation of the polypeptide.
Finally, it completes the mechano-chemical cycle by translocating itself 
completely to the next codon and is ready to begin the next cycle. 
Elongation factors (EF), which are themselves proteins, play important 
roles in the control of these major steps (see the fig.\ref{ribocartoon}) 
which require proper communication and coordination between the two 
subunits. Moreover, some specific steps in the mechano-chemical cycle 
of a ribosome are driven by the hydrolysis of guanosine triphosphate 
(GTP) to guanosine diphosphate (GDP). 

The detailed chemo-mechanical cycle of a ribosome is drawn schematically 
in fig.\ref{states}. Let us begin the biochemical cycle of a ribosome 
in the elongation phase with state 1, (figure \ref{states}) where a 
tRNA is bound to the site P. A tRNA-EF-Tu complex (a macromolecular 
complex of a tRNA and an elongation factor Tu) now binds to site A and,  
after the correct recognition of the cognate aa-tRNA through proper 
codon-anticodon matching, the system makes a transition from the state 
1 to the state 2. As long as the EF-Tu is attached to the tRNA, 
codon-anticodon binding can take place, but the peptide bond formation 
is not possible. The EF-Tu has a GTP part which is hydrolyzed to GDP, 
driving the transition from state 2 to 3. Following this, a phosphate 
group leaves, resulting in the intermediate state 4. This hydrolysis, 
finally, releases the EF-Tu, and then the peptide bond formation 
becomes possible. After this step, another elongation factor, namely, 
EF-G, in the GTP bound form, comes in contact with the ribosome. This 
causes the tRNAs to shift from site P to E and from site A to P, site 
A being occupied by the EF-G, resulting in the state 5. Hydrolysis of 
the GTP to GDP then releases the EF-G and this is accompanied by the 
transition of the system from the state 5 to the state 6. The transition 
from the state 6 to the state 7 is accompanied by conformational changes 
which are responsible for the forward movement of the smaller subunit 
by one step. Finally, the tRNA on site A is released, resulting in the 
completion of one biochemical cycle; in the process the ribosome completes 
its forward movement by one codon (i.e., one step on the lattice). 
In each cycle of the ribosome during elongation, the search for the 
cognate tRNA is the rate limiting step \cite{varenne,nierhaus}, the 
corresponding rate constant being $\omega_{a}$.

\section{Brief review of the earlier models}\label{oldmodels}

To our knowledge, MacDonald, Gibbs and coworkers 
\cite{macdonald68,macdonald69} developed the first quantitative theory 
of simultaneous protein synthesis by many ribosomes on the same mRNA 
strand. The sequence of codons on a given mRNA was represented 
by the corresponding sequence of the equispaced sites of a regular 
one-dimensional lattice. The details of molecular composition and 
architecture of the ribosomes was ignored in this model. Instead, 
each of the ribosomes was modelled by a ``self-propelled particle'' 
of size ${\ell}$ in units of the lattice constant; thus, 
${\ell}$ is an integer. On the lattice the steric interaction of 
the ribosomes was taken into account by imposing the condition of 
mutual exclusion, i.e., no site of the lattice can be simultaneously 
covered by more than one particle.

The dynamics of the system was formulated in terms of the following 
update rules:\\ 
An extended particle (effectively, a {\it hard rod}), whose forward 
edge is located at the site $i$, can hop forward by one lattice 
spacing with the forward hopping rate $q^{(i)}$ provided the target 
site is not already covered another extended particle. Moreover, 
initiation and termination were assumed to take place with the 
corresponding rates $\alpha$ and $\beta$, respectively, which are 
not necessarily equal to any of the other rate constants. 

For the sake of simplicity of analytical calculations, one usually 
replaces this intrinsically inhomogeneous process by a hypothetical 
homogeneous one by assuming \cite{macdonald68,macdonald69} that 
$q^{i} = q$, irrespective of $i$. In such special situations, this 
model reduced to the totally asymmetric simple exclusion process 
(TASEP) without any defect or disorder \cite{schuetz}, except that 
the allowed size of the ``extended'' hopping particles is a multiple 
of the lattice spacing. TASEP is one of the simplest models of 
systems of interacting driven particles \cite{sz}. Note 
that these TASEP-like earlier models of ribosome traffic 
\cite{macdonald68,macdonald69,hiernaux,gordon,vassert,vonheijne,bergmann,harley,gilchrist} capture the effects of all the chemical reactions and 
conformational changes, which lead to the translocation of a ribosome 
from one codon on the mRNA to the next, by the single parameter $q$.

The steady-state flux $J$ of the ribosomes is defined as the average 
number of the ribosomes crossing a specific codon (selected arbitrarily) 
per unit time. Because of the close analogy with vehicular traffic, 
we shall refer to the flux-density relation as the fundamental diagram 
\cite{css}. In the context of ribosomal traffic, the position, average 
speed and flux of ribosomes have interesting interpretations in terms 
of protein synthesis. The position of a ribosome on the mRNA also 
gives the length of the nascent polypeptide it has already synthesized. 
The average speed of a ribosome is also a measure of the average rate 
of polypeptide elongation. The flux of the ribosomes gives the total 
rate of polypeptide synthesis from the mRNA strand, i.e., the number of 
completely synthesized polypeptides per unit time.

The rate of protein synthesis and the ribosome density profile in the 
model developed by Macdonald et al.\cite{macdonald68,macdonald69} as 
well as in some other closely related models have been investigated 
in detail. PBC are less realistic than OBC for capturing protein 
synthesis by a theoretical model. Nevertheless, if one imposes PBC on 
this simplified version of the model, the steady-state flux of the 
ribosomes is given by \cite{macdonald68,lodish}
\begin{equation}
J = q \biggl[\frac{\rho (1-\rho~{\ell})}{1-\rho({\ell}-1)}\biggr]
\label{macgibbs}
\end{equation}
where $\rho$ is the number density of the ribosomes; if $N$ is the 
total number of ribosomes on the lattice of length $L$, then 
$\rho = N/L$. The corresponding average speed of the ribosomes is 
given by $<v> = J/\rho$. In the special case $\ell = 1$ the 
expression (\ref{macgibbs}) reduces to the well known formula
\begin{equation}
J = q ~\rho (1-\rho)
\label{fluxtasep}
\end{equation}
for the steady-state flux in the TASEP. Comparing equation 
(\ref{macgibbs}) with (\ref{fluxtasep}), $\rho/[1-\rho({\ell}-1)]$ 
has often been identified \cite{ferreira} as an {\it effective} 
particle density while the corresponding {\it effective} hole 
density is given by $1-\rho {\ell}$. The corresponding 
phenomenological hydrodynamic theory \cite{shaw03} has also been 
derived \cite{schonherr1,schonherr2} from the TASEP-like dynamics 
of the hard rods of size ${\ell}$ on the discrete lattice.

The fundamental diagram implied by the expression (\ref{macgibbs}) 
exhibits a {\it maximum} at the density $\rho_{m}$ and the value of 
flux at this maximum is $J_{m}$ where 
\begin{equation}
\rho_{m} = \frac{1}{\sqrt{\ell}(\sqrt{\ell}+1)}~~ {\rm and}~~  
J_{m} = \frac{q}{(\sqrt{{\ell}}+1)^{2}}.
\label{eq-jmax}
\end{equation} 
Only in the special case ${\ell} = 1$, this fundamental diagram is 
symmetric about $\rho = 1/2$; the maximum shifts to 
higher density with increasing ${\ell}$.

\section{The model} \label{model}

\begin{figure}[h]
\begin{center}
\includegraphics[width=0.28\columnwidth]{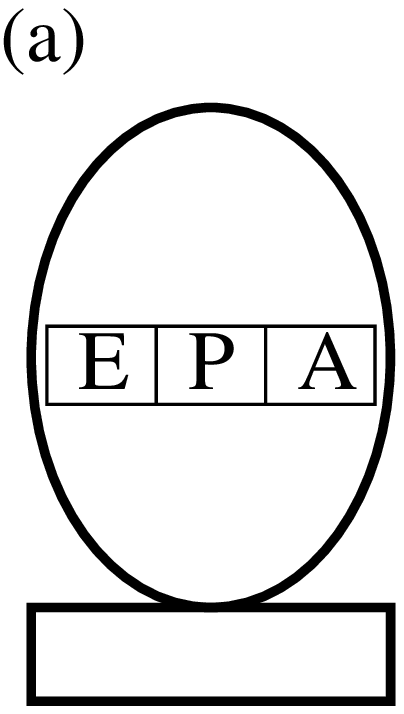}
\includegraphics[width=0.9\columnwidth]{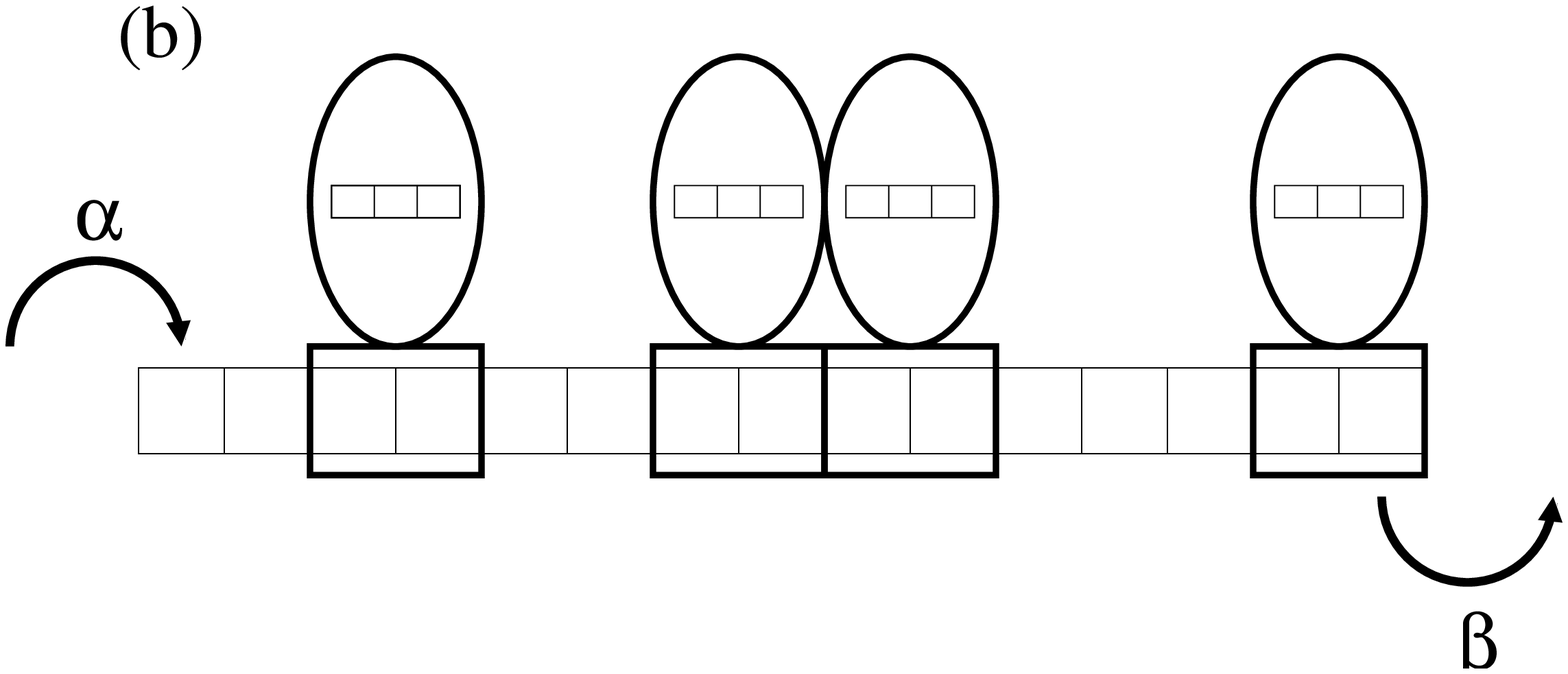}
\end{center}
\caption{A schematic representation of the model. (a) A cartoon 
of a single ribosome that explicitly shows the three binding 
sites E, P and A on the larger subunit which is represented by 
the ellipsoidal lobe. The rectangular lower part represents the 
smaller subunit of the ribosome. (b) The mRNA is represented as 
a one-dimensional lattice where each site  corresponds to one 
single codon. The smaller subunit of each ribosome covers 
${\ell}$ codons (${\ell} = 2$ in this figure) at a time. 
}
\label{fig-model}
\end{figure}

Our model is shown schematically in fig.\ref{fig-model}. 
We represent the single-stranded mRNA chain, consisting of $L$ codons, 
by a one-dimensional lattice of length $L +{\ell}-1$ where each of the 
first $L$ sites from the left represents a single codon. We label the 
sites of the lattice by the integer index $i$; the site $i = 1$ 
represents the start codon while the site $i = L$ corresponds to the 
stop codon. 

Our model differs from all earlier models in the way we capture the 
structure, biochemical cycle and translocation of each ribosome.  
The small sub-unit of the ribosome, which is known to bind to the mRNA, 
is represented by an {\it extended particle} of length ${\ell}$ which is 
expressed in the units of the size of a codon (see fig.\ref{fig-model}). 
Thus, in our model, the small subunit of each ribosome covers ${\ell}$ 
codons at a time; no lattice site is allowed to be covered simultaneously 
by more than one overlapping ribosome. Irrespective of the length $\ell$,  
each ribosome moves forward by only one site in each step as it must 
translate successive codons one by one.

Since our model is {\it not} intended to describe initiation and 
termination in detail, we represent these two processes by the 
parameters $\alpha$ and $\beta$ respectively. Whenever the first 
${\ell}$ sites on the mRNA are vacant this group of sites is allowed 
to be covered by a fresh ribosome with the probability $\alpha$ in 
the time interval $\Delta t$ (in all our numerical calculations we 
take $\Delta t = 0.001$ s). Similarly, whenever the rightmost 
${\ell}$ sites of the mRNA lattice are covered by a ribosome, i.e., 
the ribosome is bound to the $L$-th codon, the ribosome gets detached 
from the mRNA with probability $\beta$ in the time interval $\Delta t$.
Since $\alpha$ is the probability of attachment in time $\Delta t$, 
the probability of attachment per unit time (which we call 
$\omega_{\alpha}$) is the solution of the equation 
$\alpha{}=1-e^{-\omega_{\alpha}\times{}\Delta t}$ 
(see appendix A for the detailed explanation). Similarly, we also 
define $\omega_{\beta}$ which is the probability of detachment per 
unit time.

In the elongation stage, we have identified {\it seven} major distinct 
states of the ribosome in each cycle which have been described in detail 
in section \ref{bio} (shown schematically in fig.\ref{states}). However, 
in setting up the equations below, we further simplify the model. 
Throughout this paper, we replace the pathway  
$5\rightarrow{}6 \rightarrow{}7 \rightarrow 1$ by an effectively direct 
single transition $5 \rightarrow{}1$, with rate constant $\omega_{h2}$ 
(shown by the dashed line in fig.\ref{states}). This simplification is 
justified by the fact that the transitions $5\rightarrow{}6$ and 
$6\rightarrow{}7$ are purely ``internal'', and do not seem to depend on 
the availability of external molecules like elongation factors, or GTP 
or aa-tRNA.

\begin{table}
\begin{tabular}{|c|c|c|c|c|c|} \hline
$\omega_a$ ($s^{-1}$)&  $\omega_g$ ($s^{-1}$)& $\omega_p$($s^{-1}$)&  $\omega_{h1}$ ($s^{-1}$) & $\omega_{h2}$ ($s^{-1}$)& $k_{2}$ ($s^{-1}$)\\\hline
25 & 25 & 0.0028 & 10 & 10 & 2.4 \\ \hline
\end{tabular}
\caption{\label{tab-1mol}{Rate constants obtained from experimental 
data for {\it E-coli} \cite{thompson,harrington}.
 }}
\label{table-parameters}
\end{table}

The typical values of the rate constants have been extracted from 
empirical data for the bacteria {\it E-coli} \cite{thompson,harrington}. 
Moreover, since there is no significant difference in the structures of 
the two elongation factors and since their binding mechanisms are also 
similar \cite{cross97}, we assume that the rate constants $\omega_{h1}$ 
and $\omega_{h2}$ are equal. The values of the rate constants used 
in our calculations are listed in table \ref{table-parameters}.

The lifetime of a typical eukaryotic mRNA is of the order of hours 
whereas the time taken to synthesize an entire protein by translating 
the mRNA is of the order of a few minutes. Consequently, most often 
protein synthesis takes place under steady-sate conditions. Therefore, 
although we shall formulate time-dependent equations for protein 
synthesis, we shall almost exclusively focus on the steady-state 
properties of these models in this paper.

Most of our analytical calculations have been performed in the 
mean-field approximation. In order to test the accuracy of the 
approximate analytical results, we have also carried out computer 
simulations of our model. Since we found very little difference 
in the results for systems size $L = 300$ and those for larger 
systems, all of our production runs were carried out using 
$L = 300$. We have used random sequential updating which corresponds 
to the master equations formulated for the analytical description. 
In each run of the computer simulations the data for the first 
{\it five million} time steps were discarded to ensure that the 
system, indeed, reached steady state. The data were collected in 
the steady state over the next {\it five millon} time steps.
Thus, each simulation run extended over a total of ten million 
time steps. For example, the average steady-state flux was obtained 
by averaging over the last five million time steps.

\section{Results under periodic boundary conditions}\label{extended}

Typically, a single ribosome itself covers about twelve codons (i.e., 
${\ell} = 12$), and interacts with others by mutual exclusion. 
{\it The position of such a ribosome will be referred to by the 
integer index of the lattice site covered by the leftmost site of the 
smaller subunit}. Main results for the special case ${\ell} = 1$ 
are given separately in the appendix B.

\subsection{Mean field analysis under periodic boundary conditions}

Let $P_{\mu}(i)$ be the probability of finding a 
ribosome at site $i$, in the chemical state $\mu$. Also, 
$P(i) = \sum_{\mu=1}^{5} P_{\mu}(i)$, is the probability of finding 
a ribosome at site $i$, in any state. Let $P(\underbar{i}|j)$ be the 
conditional probability that, given a ribosome at site $i$, there is 
another ribosome at site $j$. Then, 
$Q(\underbar{i}|j) = 1 - P(\underbar{i}|j)$ is the conditional 
probability that, given a ribosome in site i, site j is empty. In 
the mean-field approximation, the Master equations for the 
probabilities $P_{\mu}(i)$ are given by
\begin{equation} \label {3:0}
\frac{\partial{}P_{1}(i)}{\partial{}t} = \omega_{h2} P_{5}(i-1) Q(\underbar{i-1}|i-1+\ell) + \omega_{p} P_2(i) - \omega_{a} P_{1}(i)
\end{equation}
\begin{equation} \label {3:3}
\frac{\partial{}P_{2}(i)}{\partial{}t} = \omega_{a} P_{1}(i) - (\omega_{p} + \omega_{h1}) P_{2}(i)
\end{equation}
\begin{equation} \label {3:4}
\frac{\partial{}P_{3}(i)}{\partial{}t} = \omega_{h1} P_{2}(i) - k_{2} P_{3}(i)
\end{equation}
\begin{equation} \label {3:5}
\frac{\partial{}P_{4}(i)}{\partial{}t} = k_{2} P_{3}(i) - \omega_{g} P_{4}(i)
\end{equation}
\begin{equation} \label {3:6}
\frac{\partial{}P_{5}(i)}{\partial{}t} = \omega_{g} P_{4}(i) - \omega_{h2} P_{5}(i) Q(\underbar{i}|i+{\ell}) 
\end{equation}
Note that not all of the five equations (\ref{3:0})-(\ref{3:6}) are
independent of each other because of the condition
\begin{equation}\label {gg}
P(i)=\sum_{\mu=1}^{5}P_{\mu}(i)=\frac{N}{L} = \rho
\end{equation}
In our calculations below, we have used the equations 
(\ref{3:3})-(\ref{gg}) as the five independent equations.
Mean field approximation has entered through our implicit assumption  
that the probability of there being a ribosome at site $i$ is not 
affected by the presence or absence of other ribosomes at other sites.

\subsection{Steady state properties under periodic boundary conditions}

In the steady state, all the $P_{\mu}(i)$ become independent of time. 
Moreover, if PBC are imposed (i.e., the lattice effectively forms a
ring), no site has any special status and the index $i$ can be dropped.
The corresponding flux of the ribosomes $J$ can then be obtained from
\begin{equation}
J = \omega_{h2}P_{5}Q(\underbar{i}|i+{\ell}).
\label{eq-fluxl}
\end{equation}
using the steady-state expressions for $Q(\underbar{i}|i+{\ell})$ 
and $P_{5}$.  

Because of the translational invariance in the steady state, we have 
$Q(\underbar{i}|j)=Q(\underbar{1}|j-i+1)$. Therefore, we now calculate 
$Q(\underbar{1}|1+{\ell})$: given a ribosome at the site $i=1$, what 
is the probability that the site $i={\ell}+1$ is empty? Since it is 
given that the site $i=1$ is occupied by a ribosome, the remaining 
$N-1$ ribosomes must be distributed among the remaining $L-{\ell}$ 
sites. Let us introduce the symbol $Z(L,N,{\ell})$ to denote the 
number of ways of arranging the $N$ ribosomes and $L-N{\ell}$ gaps. 
Obviously,
\begin{equation} \label{3:1}
Z(L,N,{\ell})=\frac{(N+L-N{\ell})!}{N!(L-N{\ell})!}
\end{equation}
In case it is given that one ribosome occupies $i=1$, the total number of 
configurations would be $Z(L-{\ell},N-1,{\ell})$. Of these, we wish to 
find the number of those configurations where $i={\ell}$ is also occupied; 
this is given by $Z(L-2{\ell},N-2,{\ell})$. Therefore, the probability 
that $i={\ell}$ is occupied, given that $i=1$ is occupied, is given by
\begin{eqnarray} \label{3:2}
P(\underbar{1}|{\ell}+1)&=&\frac{Z(L-2{\ell},N-2,{\ell})}{Z(L-{\ell},N-1,{\ell})} \nonumber\\
&=&\frac{N-1}{L+N-N{\ell}-1}
\end{eqnarray}
Hence,
\begin{eqnarray} \label{3:2}
Q(\underbar{i}|i+{\ell})= \frac{L-N{\ell}}{L+N-N{\ell}-1}.
\label{eq-qpbc}
\end{eqnarray}

Solving the equations (\ref{3:3}-\ref{gg}) in the steady state under 
PBC, we get 
\begin{equation}
P_{5} = \frac{P}{1+\frac{\Omega_{h2}(L - N{\ell})}{L+N-N{\ell}-1}}
\label{eq-p5}
\end{equation}
where,
\begin{equation}
\Omega_{h2} = \omega_{h2}/k_{eff}.
\end{equation}
with
\begin{equation}
\frac{1}{k_{eff}} = \frac{1}{\omega_{g}}+\frac{1}{k_{2}}+\frac{1}{\omega_{h1}}+\frac{1}{\omega_{a}}+\frac{\omega_{p}}{\omega_{a}\omega_{h1}}
\end{equation}
Note that $k_{eff}^{-1}$ is an effective time that incorporates the
delays induced by the intermediate biochemical steps in between two
successive hoppings of the ribosome from one codon to the next. 
Therefore, $k_{eff} \rightarrow \infty$ implies short-circuiting 
the entire biochemical pathway so that a newly arrived ribosome at 
a given site is instantaneously ready for hopping onto the next site 
with the effective rate constant $\omega_{h2}$.

Using expressions (\ref{3:2}) and (\ref{eq-p5}) in (\ref{eq-fluxl}) and 
the definition $\rho = N/L$ for the number density, we get 
\begin{eqnarray} \label{3:7}
J = \frac{\omega_{h2} \rho (1- \rho {\ell})}{(1+\rho-\rho {\ell})+ \Omega_{h2}(1-\rho {\ell})}
\end{eqnarray}
Note that $J$ vanishes at $\rho = 0$ and for all 
$\rho \geq \rho_{max} = 1/{\ell}$ because at the density $\rho_{max}$ 
the entire mRNA in fully covered by ribosomes. Moreover, in the special 
situation where $k_{eff} \rightarrow \infty$, but $\omega_{h2} = q$ 
remains non-zero and finite, $\Omega_{h2} \rightarrow 0$ and the 
expression (\ref{3:7}) reduces to the expression (\ref{macgibbs}). 

\begin{figure} 
\begin{center}
\includegraphics[width=0.9\columnwidth]{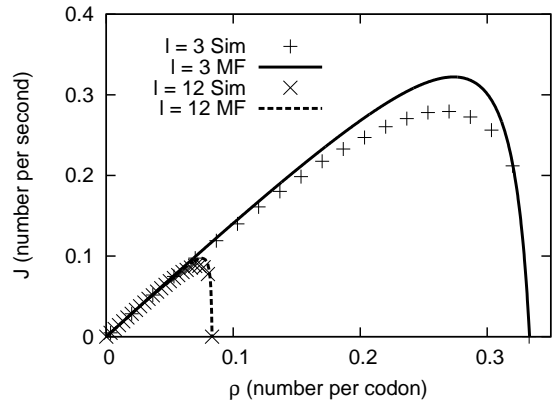}
\end{center}
\caption{Flux of ribosomes with ${\ell} = 3, 12$, under periodic 
boundary conditions, plotted against density. The curves correspond 
to the analytical expressions obtained in the mean-field (MF) 
approximation whereas the discrete data points have been obtained 
by carrying out computer simulations (Sim). Values of all the 
parameters, except $\ell$, are same as those listed in table I.
}    
\label{figa}
\end{figure}

The flux obtained from (\ref{3:7}) has been plotted in figure (\ref{figa}) 
for two values of ${\ell}$. This trend of variation with ${\ell}$ 
was also observed in the pioneering work of MacDonald et al.
\cite{macdonald68} in their simpler model. 
By differentiating equation (\ref{3:7}), we obtain that the density
$\rho^{*}$ corresponding to the maximum of the flux is the solution 
of the equation:
\begin{equation} \label{4:1}
\rho^{2}{\ell}(1-{\ell}-\Omega_{h2}{\ell})+2\rho{}{\ell}(1+\Omega_{h2})-(1+\Omega_{h2})=0
\end{equation}
and, hence, 
\begin{eqnarray}
\rho^{*} = \sqrt{\biggl(\frac{1+\Omega_{h2}}{\ell}\biggr)} \biggl[\frac{1}{\sqrt{{\ell}(1+\Omega_{h2})}+1}\biggr]
\label{soln2}
\end{eqnarray}
We recover equation (\ref{eq-jmax}) for $\rho_{m}$ in the 
appropriate limit $\Omega_{h2} \rightarrow 0$. Our theoretical 
predictions in fig.\ref{figa} are also in good agreement with 
the corresponding simulation data. 
 
\begin{figure} 
\begin{center}
\includegraphics[width=0.9\columnwidth]{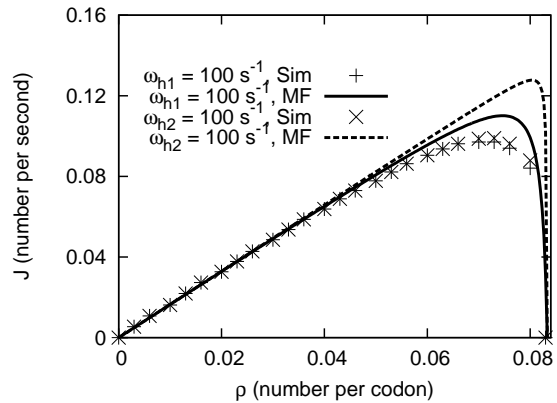}
\end{center}
\caption{Flux of ribosomes with ${\ell} = 12$, under periodic
boundary conditions, plotted against density. The curves correspond
to the analytical expressions obtained in the mean-field (MF) 
approximation whereas the discrete data points have been obtained 
by carrying out computer simulations (Sim). Values of all the 
parameters, which are not shown explicitly on the figure, 
are identical to those in table I.
}    
\label{fig-fundaomh}
\end{figure}

In order to see the effects of varying the rates of some of the
biochemical transitions, we have plotted the fundamental diagram
in fig.\ref{fig-fundaomh} for two different situations, namely,
$\omega_{h1} = 10 \omega_{h2}$ with $\omega_{h2} = 10$ s$^{-1}$
and $\omega_{h2} = 10 \omega_{h1}$ with $\omega_{h1} = 10$ s$^{-1}$.
The fundamental diagrams in these two situations turn out to be
almost identical; this is a consequence of the fact that for the
set of parameter rages used in this figure, neither $\omega_{h1}$
nor $\omega_{h2}$ corresponds to the rate limiting process.

\begin{figure}
\begin{center}
\includegraphics[width=0.9\columnwidth]{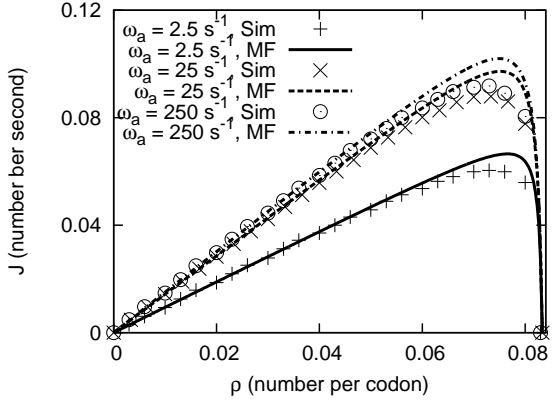}
\end{center}
\caption{Same as in fig.\ref{fig-fundaomh} except that different curves 
correspond to different values of $\omega_{a}$. Values of all the
parameters, which are not shown explicitly on the figure,
are identical to those in table I.
}
\label{fig-fundaoma}
\end{figure}

\begin{figure}
\begin{center}
\includegraphics[width=0.9\columnwidth]{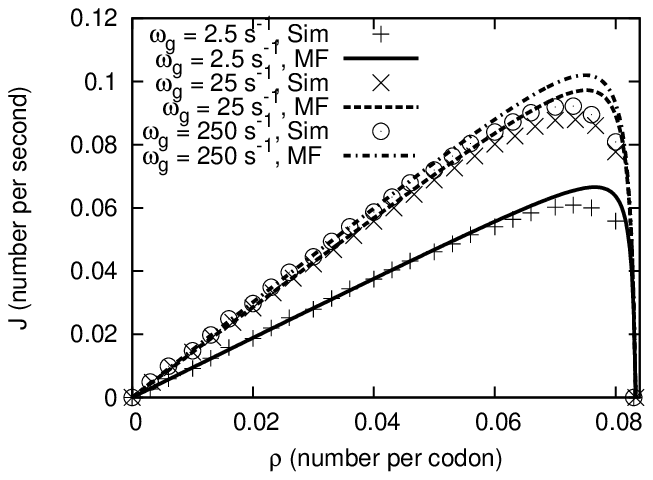}
\end{center}
\caption{Same as in fig.\ref{fig-fundaomh} except that different curves
correspond to different values of $\omega_{g}$. Values of all the
parameters, which are not shown explicitly on the figure,
are identical to those in table I.
}
\label{fig-fundaomg}
\end{figure}

\begin{figure}
\begin{center}
\includegraphics[width=0.9\columnwidth]{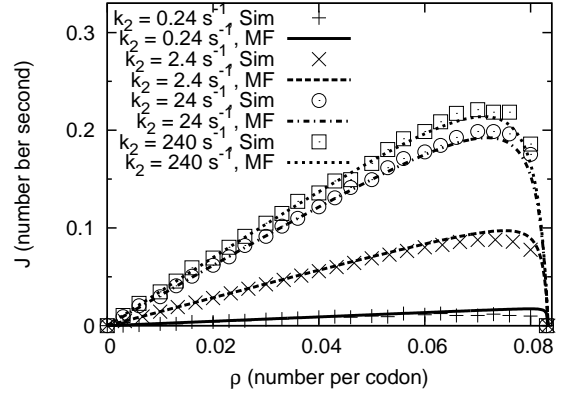}
\end{center}
\caption{ Same as in fig.\ref{fig-fundaomh} except that different curves
correspond to different values of $k_2$. Values of all the
parameters, which are not shown explicitly on the figure,
are identical to those in table I.
}
\label{fig-fundak2}
\end{figure}

We have plotted the fundamental diagrams of the model in
fig.\ref{fig-fundaoma} for three different different values of
$\omega_{a}$, namely, $\omega_{a} = 2.5$ s$^{-1}$,
$\omega_{a} = 25$ s$^{-1}$ and  $\omega_{a} = 250$ s$^{-1}$ using
both mean-field theory and computer simulations. The results show
that at sufficiently small values of $\omega_{a}$, where the
availability of the tRNA is the rate-limiting process, the flux
increases rapidly with increasing $\omega_{a}$. However, the rate
of this increase decreases with increasing $\omega_{a}$ and,
eventually, the flux essentially saturates. This saturation of 
flux occurs when $\omega_{a}$ is so large that the availability 
of tRNA is no longer the rate limiting process. A similar trend 
of variation of flux with $\omega_{g}$ is observed in 
fig.\ref{fig-fundaomg} when $\omega_{g}$ is varied over three orders 
of magnitude. In contrast, the flux has been observed to vary at a 
significant rate even at the highest values of $k_{2}$, when it is 
varied over three orders of magnitude (see fig.\ref{fig-fundak2} 
indicating that saturation of flux with respect to $k_{2}$ variation 
sets in at even higher values of $k_{2}$).
 
\section{Results under open boundary conditions}\label{openbc}

An OBC is more realistic than a PBC for describing ribosome traffic 
on mRNA. The parameters $\alpha$ and $\beta$, which are associated,  
respectively, with initiation and termination of translation play 
significant roles in the system under OBC.

\subsection{Mean field analysis under open boundary conditions}

In this subsection we calculate the flux of ribosomes (and, hence, 
the rate of protein synthesis) using a 
mean field theoretical approach similar to that developed by Shaw et 
al. \cite{shaw04a}. The approximation involved is that the conditional 
probability of site $i+{\ell}$ being empty, given that site $i$ has a 
ribosome in it, is replaced simply by the probability of site $i$ 
being empty, given no other condition. If $P(i)$ is the probability 
of there being a ribosome at site $i$, then the probability of there 
being a hole at site $j$ is given by $1-\sum_{s=0}^{{\ell}-1}P(j-s)$. 
It is now straightforward to set up the master equations for the 
probabilities $P_{\mu}(i)$ in the mean-field approximation:
\begin{equation} \label{a}
\frac{dP_{1}(1)}{dt}=\omega_{\alpha}(1-\sum_{s=1}^{{\ell}}P(s))+\omega_{p}P_{2}(1)-\omega_{a}P_{1}(1)
\end{equation}
\begin{eqnarray} \label{b}
\frac{dP_{1}(i)}{dt}=\frac{\omega_{h2}P_{5}(i-1)(1-\sum_{s=1}^{{\ell}}P(i-1+s))}{1-\sum_{s=1}^{l}P(i-1+s)+P(i-1+{\ell})} \nonumber \\
+\omega_{p}P_{2}(i)-\omega_{a}P_{1}(i)\\
(i\ne{}1)\nonumber
\end{eqnarray}
\begin{equation} \label{c}
\frac{dP_{2}(i)}{dt}=\omega_{a}P_{1}(i)-(\omega_{p}+\omega_{h1})P_{2}(i)
\end{equation}
\begin{equation} \label{d}
\frac{dP_{3}(i)}{dt}=\omega_{h1}P_{2}(i)-k_{2}P_{3}(i)
\end{equation}
\begin{equation} \label{e}
\frac{dP_{4}(i)}{dt}=k_{2}P_{3}(i)-\omega_{g}P_{4}(i)
\end{equation}
\begin{eqnarray} \label{f}
\frac{dP_{5}(i)}{dt}=\omega_{g}P_{4}(i)-\frac{\omega_{h2}P_{5}(i)(1-\sum_{s=1}^{{\ell}}P(i+s))}{1-\sum_{s=1}^{{\ell}}P(i+s)+P(i+{\ell})}\\
(i\ne{}N)\nonumber
\end{eqnarray}
\begin{equation} \label{g}
\frac{dP_{5}(N)}{dt}=\omega_{g}P_{4}(N)-\omega_{\beta} P_{5}(N)
\end{equation}

\subsection{Steady state properties under open boundary conditions}

The flux is given by $J=\omega_{\alpha}(1-\sum_{s=0}^{{\ell}}P(s))$. 
This flux has been computed numerically by solving 
equations (\ref{a}-\ref{g}); the results are shown by the continuous 
curves in figures \ref{fig9}(a) and \ref{fig9}(b). These mean-field 
estimates are in excellent agreement with the corresponding numerical 
data obtained from computer simulations of the model. Moreover, the 
rates of protein synthesis corresponding to the typical rate constants 
given in table I are in the same order of magnitude as those observed 
experimentally \cite{spirinbook}. The average density profiles observed at 
several values of $\omega_{\alpha}$ and $\omega_{h}$ are also shown 
in the insets of figs.\ref{fig9}(a) and (b), respectively.

Figures \ref{fig9}(a) and (b) show how the current gradually increases 
and, finally, saturates as $\omega_{\alpha}$ (in (a)) and $\omega_h$ 
(in (b))  increases; the saturation value of the current is numerically 
equal to the maximum current obtained in the corresponding closed 
system with periodic boundary conditions. The average density profiles 
in the insets of the figures \ref{fig9}(a) and (b) establish that the 
average density of the ribosomes {\it increases} with increasing 
$\omega_{\alpha}$, but {\it decreases} with increasing $\omega_{h}$, 
gradually saturating in both the cases. These observations are 
consistent with the scenario of phase transition from one dynamical 
phase to another, as predicted by the extremal current hypothesis 
which will be considered later in this section.

\subsection{Effects of inhomogeneity of the mRNA track of real gene sequences}

In a real mRNA the nucleotide sequence is, in general, inhomogeneous. 
First of all, different codons appear on an mRNA with different 
frequencies. Moreover, in a given cell, not all the tRNA species, 
which correspond to different codon species, are equally abundant. 
Interestingly, because of evolutionary adaptations, the concentrations 
of tRNA species which correspond to rare codons are also proportionately 
low \cite{andersson90}. Thus, sequence inhomogeneity on a real mRNA 
can have important effects on the speed and fidelity of translation 
\cite{bergmann,harley,chou04,zhang}. 

We now extend our homogeneous model {\it assuming} that the rate 
constant $\omega_{a}$ of the attachment of the tRNA to the site $A$ 
of the ribosome is site-dependent (i.e., dependent on the codon 
species). More precisely, for a ribosome located at the $i$-th site, 
we multiply the numerical value of $\omega_{a}$, which we used 
earlier for the hypothetical homogeneous mRNA, by a multiplicative 
factor that is proportional to the relative concentration of the 
tRNA associated with the $i$-th codon \cite{andersson90,solomo}.

\begin{figure}
\begin{center}
\includegraphics[width=0.9\columnwidth]{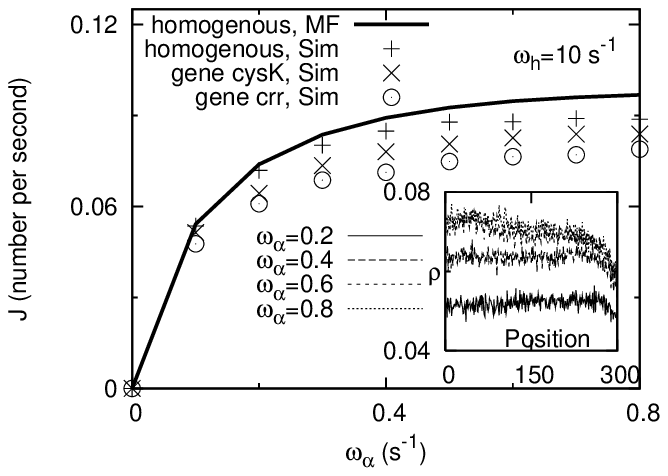}
(a)
\includegraphics[width=0.9\columnwidth]{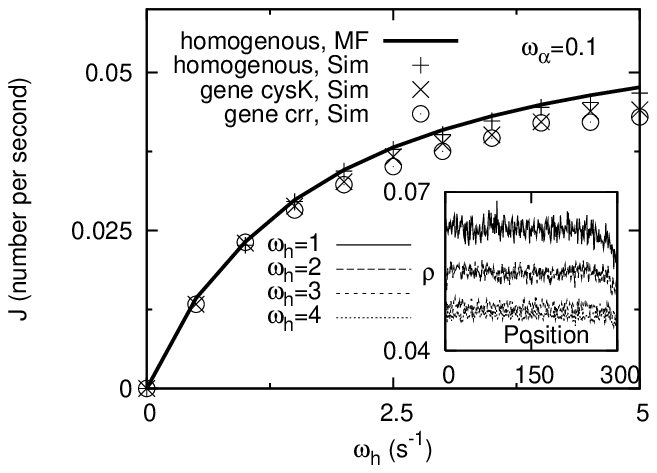}
(b)
\end{center}
\caption{Flux of ribosomes plotted against (a) $\omega_{\alpha}$ and 
(b) $\omega_h$ for the genes  crr (170 codons) and cysK (324 codons)
of $Escherichia~coli$ K-12 strain MG1655, as well as the
corresponding curve for a homogenous mRNA strand of 300 codons.
The insets show the average density profiles on a hypothetical
{\it homogeneous} mRNA track for four different values of
(a) $\omega_{\alpha}$ and (b) $\omega_h$, for fixed 
$\omega_a = 25$ s$^{-1}$.
}
\label{fig9}
\end{figure}

A lot of work on TASEP with quenched random space-dependent hopping rates 
\cite{lebowitz,schutzdis,tripathi,goldstein,kolwankar,harris,derrida,santen2,barmarev} 
and Brownian motors with quenched disorder 
\cite{harms,marchesoni,family,jia} has been reported. Similarly, 
effects of randomness on some stochastic chemical kinetic models of 
molecular motors have also been investigated \cite{nelson1,nelson2}. 
However, the nucleotide sequence on any real DNA or mRNA is not random.
But, to our knowledge, for the inhomogeneous, but correlated, gene 
sequences no analytical technique is available at present for the 
calculation of the speed of the associated molecular motors. For 
example, all the theoretical schemes developed so far for RNA 
polymerase motors \cite{wang,ruckenstein} by taking into account the 
actual sequence of the specific DNA track, have to be implemented 
numerically. Even in the context of earlier TASEP-like models of 
protein synthesis, almost all the theoretical results on the effects 
of sequence inhomogeneities have been obtained by computer simulations 
\cite{dong}. Therefore, we have been able to study the effects of 
sequence inhomogeneities of real codon sequences on the rate of protein 
synthesis in our model only numerically by carrying out computer 
simulations.

In our numerical studies, we focus on genes of {\it Escherichia~coli} 
K-12 strain MG1655 \cite{databank}. We use the hypothesis mentioned 
above for the choice of the numerical values for the different species 
of codons. The results of our computer simulations are plotted with 
discrete points in fig.{\ref{fig9}. The lower flux observed for real 
genes, as compared to that for a homogeneous mRNA, is caused by the 
codon specificity of the available tRNA molecules.

\subsection{Phase diagrams from extremal current hypothesis}

We shall treat $\alpha$, $\omega_{a}$, $\omega_{h1}$ and $\omega_{h2}$ 
as the experimentally controllable parameters. We shall plot the phase 
diagrams of the model in planes spanned by pairs of these parameters.
We shall plot these phase diagrams using equation (\ref{3:7}), and an 
extremum principle which was originally introduced by Krug \cite{krug91}, 
stated in its general form by Popkov and Sch\"utz \cite{popkov2} and 
effectively utilized in several later works \cite{hager1,hager2} in 
the context of driven diffusive lattice gas models.

\begin{figure} 
\begin{center}
\includegraphics[angle=-90,width=0.9\columnwidth]{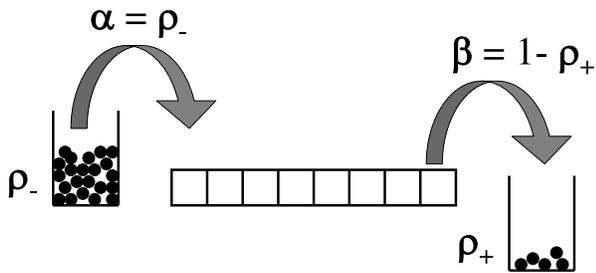}
\end{center}
\caption{Incorporating $\alpha$ and $\beta$ through two reservoirs 
with appropriate densities.}
\label{fig-obcextrem}
\end{figure}

In this approach, one imagines that the left and right ends of the 
system are connected to two reservoirs with the appropriate number 
densities $\rho_{-}$ and $\rho_{+}$ of particles (ribosomes) so 
that, assuming the same jumping rates as in the bulk, the rates 
$\alpha$ and $\beta$ are incorporated into the model 
(see fig.\ref{fig-obcextrem}). 

The extrema principle then relates the flux ${\cal J}$ in the open 
system to the flux $J(\rho)$ for the corresponding closed system 
(i.e., the system with periodic boundary conditions) with the same 
dynamics. In the limit $L\to{}\infty$ \cite{popkov2}, the extrema 
principle states that 
\begin{eqnarray}
{\cal J} = 
\left\{
\begin{array}{ccc}
max~J(\rho) & {\rm if}~\rho_{-} > \rho > \rho_{+}\\
min~J(\rho) & {\rm if}~\rho_{-} < \rho < \rho_{+}
\end{array} \right.
\end{eqnarray}

In the present context of our model the expression (\ref{3:7}) for 
$J(\rho)$ exhibits a single maximum at $\rho = \rho^{*}$ where 
$\rho^{*}$ is given by the equation (\ref{soln2}). Moreover, we 
take $\rho_{+}=0$, i.e., $\beta=1$, because we assume that the 
ribosome is released from the mRNA as soon as it reaches the stop 
codon; this is justified by the fact that, normally, termination is 
not the rate limiting step in the process of protein synthesis. 
Therefore, the extremal current hypothesis implies that in our model 
\begin{equation}
{\cal J} = max~J(\rho)  ~{\rm if}~\rho_{-} > \rho^{*} 
\label{eq-exsingle}
\end{equation}
All the results derived in this section exploiting this extremum 
principle are approximate because the expression (\ref{3:7}), which 
we use for the expression of $J(\rho)$, has been derived in the 
mean-field approximation. Next, following arguments similar to those 
followed in all earlier applications of this extremum principle, 
we now derive the appropriate expressions for $\rho_{-}$.

Consider a closed system with $L$ sites. Given that a sequence of 
${\ell}$ successive sites are empty, the total number of ways in 
which $N$ ribosomes can be distributed over the remaining $L-{\ell}$ 
sites is simply $Z(L-{\ell},N,{\ell})$. Of these, 
$Z(L-2{\ell},N-1,{\ell})$ configuations have a ribosome in the 
adjacent ${\ell}$ sites to the {\it left} of these empty ${\ell}$ 
sites. Let us use the symbol 
${\cal P}({\underbrace{11...1}_{\ell}}|{\underbrace{\underbar{00...0}}_{\ell}})$ 
for the conditional probability that, given a sequence of ${\ell}$ 
successive empty sites, there will be a ribosome in the adjacent 
${\ell}$ sites to its {\it left}. Then, from the above considerations, 
\begin{equation}
{\cal P}({\underbrace{11...1}_{\ell}}|{\underbrace{\underbar{00...0}}_{\ell}})
= \frac{Z(L-2{\ell},N-1,{\ell})}{Z(L-{\ell},N,{\ell})} = \frac{N}{L-{\ell}+N-N{\ell}}. 
\end{equation}
In terms of the number density $\rho$, this probability can expressed 
as 
\begin{equation}
{\cal P}({\underbrace{11...1}_{\ell}}|{\underbrace{\underbar{00...0}}_{\ell}})
= \frac{\rho}{\rho{}+1-\frac{{\ell}}{L}-\rho{}{\ell}}. 
\label{eq-undbrp}
\end{equation}
Moreover, solving the equations (\ref{a}-\ref{g}) we find 
\begin{equation} \label {4:7}
P_{5}=\frac{1}{1+\Omega_{h2}}
\end{equation}
Therefore, if $P^{jump}$ is the probability that, given a sequence of 
${\ell}$ successive empty sites, a ribosome will hop onto it in the 
next time step, we have 
\begin{eqnarray} \label{4:8}
P^{jump}&=&{\cal P}({\underbrace{11...1}_{\ell}}|{\underbrace{\underbar{00...0}}_{\ell}}) \times{}P_{5}\times{}\omega_{h2} (\Delta t)
\end{eqnarray}
where 
${\cal P}({\underbrace{11...1}_{\ell}}|{\underbrace{\underbar{00...0}}_{\ell}})$
and $P_{5}$ are given by (\ref{eq-undbrp}) and (\ref{4:7}), 
respectively.

Now, going back to the open system, $\rho_{-}$ is the solution of the 
equation $\alpha{}=P^{jump}$ and, hence, we get 
\begin{equation} \label{4:9}
\rho_{-}=\frac{\alpha{}(1-\frac{{\ell}}{L})(1+\Omega_{h2})}{P_{\omega_h} -\alpha
{}(1+\Omega_{h2})(1-{\ell})}
\end{equation}
where $P_{\omega_{h}}$ is the probability of hydrolysis in 
time $\Delta{}t$. In the special case $k_{eff} \rightarrow \infty$ 
while $\omega_{h2} = q$ remains finite and non-zero, i.e., 
$\Omega_{h2} \rightarrow 0$, 
$\rho_{-} \rightarrow \frac{\alpha}{1+({\ell}-1)\alpha}$, 
which is identical to the expression derived earlier by Lakatos and 
Chou \cite{lakatos03} for this special case.

\begin{figure} 
\begin{center}
\includegraphics[width=0.9\columnwidth]{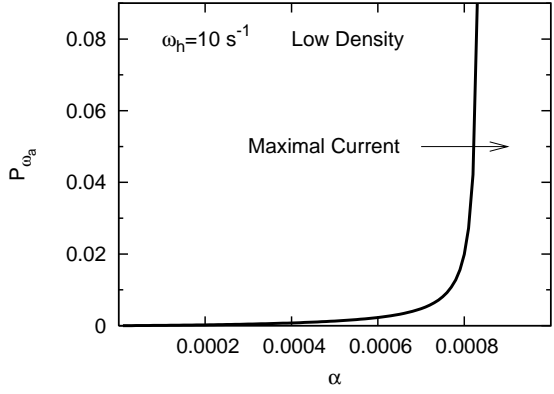}
\end{center}
\caption{Phase diagram in $\alpha{} - P_{\omega_{a}}$ plane. 
$P_{\omega_{a}}$ is the probability of attachment of a tRNA in time 
$\Delta{}t=0.001~s$, and is related to $\omega_{a}$ by equation 
(\ref{rate2}). This diagram has been plotted for 
$\omega_{h1} = \omega_{h2} = \omega_{h} = 10~sec^{-1}$. 
}    
\label{fig6}
\end{figure}

\begin{figure} 
\begin{center}
\includegraphics[width=0.9\columnwidth]{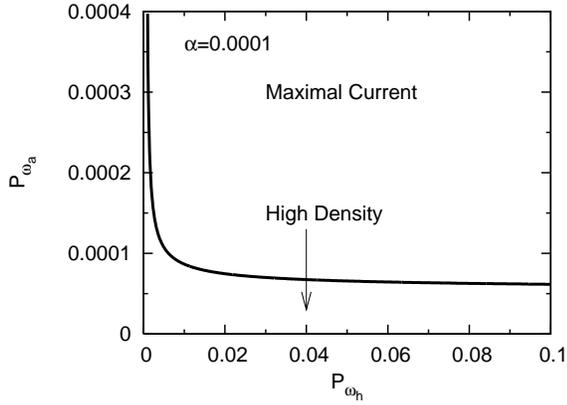} 
\end{center}
\caption{Phase diagram in $P_{\omega_{h}} - P_{\omega_{a}}$ plane. 
$P_{\omega_{h}}$ is the probability of hydrolysis in time 
$\Delta{}t=0.001~s$, and is related to $\omega_{h}$ by equation 
(\ref{rate2}).  
}    
\label{fig7}
\end{figure}

Thus, the boundaries between various phases have been obtained by 
solving the equation
\begin{equation} \label{4:10}
\rho_{-}(\alpha{},\omega_{a},\omega_{h1},\omega_{h2})=\rho_{*}(\alpha{},\omega_{
a},\omega_{h1},\omega_{h2})
\end{equation}
numerically using $\rho^{*}$ and $\rho_{-}$ obtained, respectively, 
from equations (\ref{soln2}) and (\ref{4:9}); two typical phase diagrams 
have been plotted in in figures \ref{fig6} and \ref{fig7} assuming 
\cite{thompson,harrington,cross97} $\omega_{h1} = \omega_{h2} = \omega_{h}$.
Each of these phase diagrams show two phases namely, a maximal current 
phase and another phase. In order to find out whether the latter is 
the low-density phase or the high-density phase, we studied the 
trend of variation of the density profile across the phase boundary 
(see fig.\ref{fig9}). If the average density {\it increases} and, 
finally, saturates in the maximal current phase (as observed in the 
inset of fig.\ref{fig9}(a)) while the current reaches its maximum 
value, we identify it with the transition from the low-density phase 
to the maximal current phase (as in fig.\ref{fig6}). On the other hand, 
gradual {\it decrease} of the average density and its eventual 
saturation (as observed in the inset of fig.\ref{fig9}(b)), while the 
current approaches its maximum value, indicates a transition from the 
high-density phase to the maximal current phase (as in fig.\ref{fig7}). 

Although the extremum current hypothesis \cite{popkov2} is believed 
to be exact, at least in the limit $L \rightarrow \infty$, our results 
on the phase boundaries are approximate because we have used the 
mean-field estimates (\ref{soln2}) and (\ref{4:9}) for $\rho^{*}$ and 
$\rho_{-}$, respectively, in equation (\ref{4:10}).

\section{Summary and conclusion}\label{conclusion}

TASEP is the simplest model of systems of interacting ``self-propelled'' 
particles. Interestingly, the TASEP itself was developed originally 
\cite{macdonald68} to describe traffic-like collective movement of 
ribosomes on an mRNA strand. The physical properties of the TASEP and 
similar driven-diffusive lattice gas models have been investigated 
extensively in the recent years using the techniques of non-equilibrium 
statistical mechanics \cite{sz,schuetz}. Success of TASEP-like models 
in  vehicular traffic \cite{css} has not only led to the recent modelling 
of  molecular motor traffic by suitable extensions of TASEP
\cite{polrev,tgf06,physica,lipo3,frey3,santen,popkov1}, 
but has also revived interest in ribosome traffic
\cite{macdonald68,macdonald69,lakatos03,shaw03,shaw04a,shaw04b,chou03,chou04}. 

In reality, a ribosome is not just a ``particle'' but one of the most 
complex natural nanomotors \cite{spirinbook}. An underlying implicit 
assumption of the TASEP type models of ribosome traffic is that the 
numerical value of the hopping rate $q$ is determined by the main 
rate-limiting step in the mechano-chemical cycle of a ribosome. 
In contrast, in this paper we have developed a model of ribosome traffic 
during protein synthesis by explicitly incorporating all the major steps 
in the mechano-chemical cycle of each ribosome, in addition to the mutual 
exclusion of the ribosomes arising from their steric interactions. 
Thus, our work can be viewed as an interesting 
biologically motivated extension of TASEP to an exclusion process for 
extended particles with ``internal states''. Exclusion processes with 
``internal states'' have begun receiving attention in very recent 
literature \cite{reichenbach,tabatabei}.

We have calculated, both analytically and numerically, the flux of 
the ribosomes, which is directly related to the rate of protein 
synthesis. We have investigated how the rate of protein synthesis 
is affected by the variation of the rate constants for the various 
steps of the mechano-chemical cycle of individual ribosomes. We have 
demonstrated that, with the increase of the numerical value of a 
rate constant, the current eventually saturates when the corresponding 
step of the mechano-chemical cycle is no longer rate limiting.

We have also calculated the average density profiles of 
the ribosomes on the mRNA track in all the dynamical phases of the 
system. Using a few real mRNA sequences for {\it E-coli}, we have 
demonstrated that, because of the sequence inhomogeneity, the rate 
of protein synthesis from real mRNA templates is slower than that 
from a hypothetical homogeneous template. Besides, we have calculated 
the flux in real time (unlike arbitrary units used in most of the 
earlier works). Our theoretical estimates for the rates of protein 
synthesis are in good agreement with the corresponding experimental data.

Our work also elucidates the nature of boundary-induced non-equilibrium 
phase transitions in a biologically-motivated driven-diffusive lattice 
gas model \cite{schuetz}. We have determined the phase boundaries 
on the phase diagrams for our model by using the extremum current 
hypothesis \cite{popkov2}. Following the traditional approach to phase 
diagrams of TASEP under OBC, all the earlier works on TASEP like models 
of ribosome traffic reported phase diagrams in the $\alpha-\beta$ 
plane. But, we have plotted the phase diagrams in planes spanned by 
experimentally accessible parameters that include the concentrations 
of aa-tRNA and GTP-bound elongation factors. We hope the predictions 
of our theoretical model will stimulate further experimental studies 
for more accurate quantitative data.

\vspace{0.5cm}

\noindent {\bf Acknowledgements}: 
One of the authors (DC) thanks Stephan Grill, Frank J\"ulicher, 
Anatoly Kolomeisky, Alex Mogilner, Katsuhiro Nishinari and Gunter 
Sch\"utz for comments on an earlier version of the manuscript and/or 
for drawing attention to some earlier relevant works. This work has 
been supported (through DC) by Council of Scientific and Industrial 
Research (CSIR), government of India. 

\vspace{1cm}

\noindent{\bf {Appendix A: Rate constant versus probability}}

Consider a chemical reaction: 
A$\longrightarrow$B, with a rate constant $k$. Thus,
\begin{equation} \label{rate1}
\frac{d[B]}{dt} = k[A] = -\frac{d[A]}{dt}
\end{equation}
Solving (\ref{rate1}) gives
\begin{equation} \label {rate2}
\frac{[A]_{\circ} - [A]}{[A]_{\circ}} = 1-e^{-k\Delta{}t}
\end{equation}
where [A]$_{\circ}$ is the concentration of A at $t=0$. The left hand 
side of (\ref{rate2}) gives the fraction of A molecules reacted in 
time $\Delta{}t$, and is thus the probability that a \emph{single} 
molecule of A will be transformed into B, in time $\Delta{}t$. If this 
time interval $\Delta{}t$ is very small, we can expand the right hand 
side of equation (\ref{rate2}). Differentiation of this with restect to 
time gives the probability of transition per unit time:
\begin{equation}\label{rate3}
\omega_{A\rightarrow{}B} = \frac{\partial{}}{\partial{}t} \lim_{t\rightarrow{}0} (1-e^{-kt}) = k
\end{equation}
If $P_{A}$ is the probability of finding the molecule in state A, then 
the final master equation, according to (\ref{rate3}) is
\begin{equation} \label{rate4}
\frac{\partial{}P_{A}}{\partial{}t} = -\omega_{A\rightarrow{}B}P_{A}
\end{equation}

\noindent{\bf {Appendix B: Results in the special case ${\ell} = 1$} \label{secl=1}}

Consider the special case ${\ell} = 1$ of our model of ribosome 
traffic (with 5 ``internal'' states for each ribosome) on a mRNA 
of $L$ codons. This model is {\it not} equivalent to the TASEP-like 
model of Lakatos et al.\cite{laka05} where particles (without 
internal states) hop on a lattice of $5L$ sites with spatially 
periodic hopping rates. Under {\it periodic boundary conditions}, 
the factor $Q(\underbar{i}|i+{\ell})$ in equations (\ref{3:0}) and 
(\ref{3:6}) reduce to the simpler form $Q(\underbar{i}|i+1) = 1-P(i+1)$. 
Thus, for ${\ell} = 1$, in the steady-state with PBC,
\begin{equation} \label{2:0:11}
P_{5} = \frac{P}{1+\Omega_{h2}(1 - P)}
\end{equation}
and, hence, 
\begin{equation}
J = \omega_{h2} P_{5} (1-P) = \frac{\omega_{h2} \rho(1-\rho)}{1+\Omega_{h2}(1-\rho)}
\label{fluxl1}
\end{equation}
In the special case $k_{eff} \rightarrow \infty$, where $\omega_{h2} = q$ 
is non-zero and finite, $\Omega_{h2} \rightarrow 0$; in that case 
the expression (\ref{fluxl1}) reduces to the corresponding formula 
(\ref{fluxtasep}) for TASEP.


\end{document}